\pdfoutput=1

\documentclass[twocolumn]{emulateapj}

\usepackage{amsmath}
\usepackage{graphicx}
\usepackage{color}
\usepackage[colorlinks]{hyperref}
\hypersetup{
    colorlinks,	
    citecolor=blue,
}

\newcommand{\be}{\begin{eqnarray}}
\newcommand{\ee}{\end{eqnarray}}

\shorttitle{A reflection model with a radial disk density profile}
\shortauthors{Abdikamalov et al.}

\begin{document}

\title{A reflection model with a radial disk density profile}

\author{Askar~B.~Abdikamalov\altaffilmark{1,2}, Dimitry~Ayzenberg\altaffilmark{3}, Cosimo~Bambi\altaffilmark{1,\dag}, Honghui~Liu\altaffilmark{1} and Ashutosh~Tripathi\altaffilmark{1}}

\altaffiltext{1}{Center for Field Theory and Particle Physics and Department of Physics, 
Fudan University, 200438 Shanghai, China. \email[\dag E-mail: ]{bambi@fudan.edu.cn}}
\altaffiltext{2}{Ulugh Beg Astronomical Institute, Tashkent 100052, Uzbekistan}
\altaffiltext{3}{Theoretical Astrophysics, Eberhard-Karls Universit\"at T\"ubingen, D-72076 T\"ubingen, Germany}

\begin{abstract}
In this paper we present {\tt relxilldgrad\_nk}, a relativistic reflection model in which the electron density of the accretion disk is allowed to have a radial power-law profile. The ionization parameter has also a non-constant radial profile and is calculated self-consistently from the electron density and the emissivity. We show the impact of the implementation of the electron density gradient in our model by analyzing a \textsl{NuSTAR} spectrum of the Galactic black hole in EXO~1846--031 during its last outburst in 2019 and a putative future observation of the same source with \textsl{Athena} and \textsl{eXTP}. For the \textsl{NuSTAR} spectrum, we find that the new model provides a better fit, but there is no significant difference in the estimation of the model parameters. For the \textsl{Athena}+\textsl{eXTP} simulation, we find that a model without a disk density profile is unsuitable to test the spacetime metric around the compact object, in the sense that modeling uncertainties can incorrectly lead to finding a non-vanishing deformation from the Kerr solution.
\end{abstract}


\section{Introduction}

Blurred reflection features observed in the X-ray spectra of Galactic black holes and active galactic nuclei are thought to be produced by illumination of a cold accretion disk by a hot corona~\citep{1989MNRAS.238..729F,2003PhR...377..389R,2013Natur.494..449R,2020arXiv201104792B}. For geometrically thin and optically thick accretion disks, the gas is in local thermal equilibrium and every point on the disk has a blackbody-like spectrum. The whole disk has a multi-temperature blackbody-like spectrum, which is normally peaked in the soft X-ray band for stellar-mass black holes and in the UV band for the supermassive ones. The ``corona'' is some hotter gas near the black hole and the inner part of the accretion disk, but its exact morphology is not yet well understood. Thermal photons of the disk can thus inverse Compton scatter off free electrons in the corona. This process generates a continuum, which can often be approximated by a power-law spectrum with an exponential high-energy cutoff. The Comptonized photons can illuminate the disk: Compton scattering and absorption followed by fluorescent emission generate the reflection spectrum.

The reflection spectrum in the rest-frame of the gas in the disk is characterized by narrow fluorescent emission lines below 7~keV, notably the iron K$\alpha$ complex at 6.4~keV for neutral or weakly ionized iron and up to 6.97~keV for H-like iron ions, and the Compton hump peaked at 20-30~keV~\citep{2005MNRAS.358..211R,2010ApJ...718..695G}. The reflection photons detected far from the source are affected by relativistic effects during their propagation in the strong gravity region around the black hole~\citep{1989MNRAS.238..729F,1991ApJ...376...90L,2010MNRAS.409.1534D,2017bhlt.book.....B}. As a result, in the observed reflection spectrum of the source the originally narrow fluorescent emission lines become broadened and skewed. Since the reflection spectrum is mainly generated from the inner part of the accretion disk, the analysis of the reflection features in the spectra of accreting black holes can be used to study the properties of the accreting material around the compact object, measure black hole spins, and even test fundamental physics in the strong gravity regime~\citep{2013mams.book.....B,2014SSRv..183..277R,2017RvMP...89b5001B,2020arXiv201104792B}.

In the past $\sim$10~years, there have been significant efforts to develop models for the analysis of the observed relativistic reflection features in the spectra of accreting black holes and to understand the systematics affecting the measurements of the properties of these systems~\citep[see, e.g.,][and references therein]{2020arXiv201104792B}. As of now, there are four main relativistic reflection models used by the X-ray astronomy community: {\tt kyn}~\citep{2004ApJS..153..205D}, {\tt reflkerr}~\citep{2008MNRAS.386..759N,2019MNRAS.485.2942N},  {\tt relxill}~\citep{2013MNRAS.430.1694D,2014ApJ...782...76G}, and the more recent {\tt reltrans}~\citep{2019MNRAS.488..324I}. All these models employ, even if at different levels, the {\tt xillver} model~\citep{2010ApJ...718..695G,2013ApJ...768..146G} for the calculation of the reflection spectrum in the rest-frame of the gas in the disk and are today the most accurate models used in the analysis of relativistic reflection features. Another popular choice is the non-relativistic reflection model {\tt reflionx}~\citep{2005MNRAS.358..211R} together with a relativistic convolution model, but in such a case it is not possible to account for the distribution of different emission angles~\citep{2014ApJ...782...76G,2020MNRAS.498.3565T}.

Despite the significant advancements, there are still important simplifications in the current relativistic reflection models, and work is in progress to remove those simplifications affecting the final accuracy of the measurement of the properties of accreting black holes. {\tt relxill\_nk}~\citep{2017ApJ...842...76B,2019ApJ...878...91A,2020ApJ...899...80A} is an extension of the {\tt relxill} model to non-Kerr spacetimes and specifically designed to test the nature of accreting black holes~\citep{2018PhRvL.120e1101C,2019ApJ...884..147Z,2019ApJ...874..135T,2019ApJ...875...56T,2021ApJ...907...31T,2021ApJ...913...79T,2021arXiv210610982T}. In this paper, we present a new flavor of {\tt relxill\_nk}. It is called {\tt relxilldgrad\_nk} and provides the possibility of modeling the electron density of the disk, $n$, with a radial power-law profile. The ionization parameter of the disk, $\xi$, defined as
\be\label{eq-xi}
\xi = \frac{4 \pi F_{\rm X}}{n} \, ,
\ee
where $F_{\rm X}$ is the X-ray flux illuminating the disk, is calculated self-consistently in {\tt relxilldgrad\_nk} from the local values of $n$ and of the emissivity $\epsilon$, which can be approximated to be proportional to $F_{\rm X}$\footnote{Assuming a point-like corona with a power-law spectrum, one finds $F_{\rm X} (r) \propto \left[ g(r) \right]^{2-\Gamma}\epsilon(r)$ (see the appendix for its derivation), where $g = E_{\rm d}/E_{\rm c}$ is the redshift experienced by photons when they travel from the corona to the disk and $\Gamma$ is the photon index of the power-law spectrum. If we do not specify the coronal geometry, we cannot calculate $g(r)$ and we can only make the approximation $F_{\rm X} (r) \propto \epsilon(r)$.}. With the exception of the last version of {\tt reltrans} \citep{2021MNRAS.507...55M} and a currently non-public version of {\tt relxill}, all other reflection models assume that the electron density is constant over the whole disk.

We note the importance of implementing an electron density gradient into our relativistic reflection models. There is indeed increasing evidence of the necessity of employing a reflection model with a radial ionization profile in the analysis of reflection features~\citep[see, e.g.,][]{2012A&A...545A.106S,2019MNRAS.485..239K,2020MNRAS.492..405S,2021PhRvD.103j3023A}. However, a model with independent emissivity and ionization profiles is not self-consistent, because the two quantities are not independent but connected by Eq.~(\ref{eq-xi}). In {\tt relxilldgrad\_nk} we have thus two parameters describing the electron density profile, namely the electron density at the inner edge of the disk and the power-law index of the density gradient, and one parameter for the ionization of the disk, which is the maximum value of the ionization parameter. The ionization profile is then calculated using Eq.~(\ref{eq-xi}). We show the importance of this implementation by analyzing a \textsl{NuSTAR} observation of the black hole binary EXO~1846--031, as an example of an observation possible today, and a simulated observation of the same source with \textsl{Athena} and \textsl{eXTP}, to understand the impact of the disk density gradient for observations with the next generation of X-ray missions.

Our paper is organized as follows. In Section~\ref{s-disk}, we describe {\tt relxilldgrad\_nk}. In Section~\ref{s-app}, we fit the \textsl{NuSTAR} spectrum of the black hole binary EXO~1846--031 with three different models to show the impact of the electron density gradient on the measurement of the properties of the system, and then we repeat the same analysis with a simulated observation of \textsl{Athena} and \textsl{eXTP}. We discuss our results in Section~\ref{s-dc}.


\section{Radial disk density profile} \label{s-disk}

Up to very recently, relativistic reflection models employed an accretion disk with constant electron density and ionization parameter. Indeed, old attempts to add a radial ionization profile showed that observational data did not require any ionization gradient (Dauser \& Garcia, private communication), which was interpreted as due to the fact that the corona illuminated a small portion of the inner part of the accretion disk which could be approximated well by a one-zone ionization. Such an approximation was questioned in \citet{2012A&A...545A.106S} and, more recently, in \citet{2019MNRAS.485..239K}. The possibility of a non-constant ionization parameter has been thus implemented in most reflection models and its relevance in spectral fitting with current data was shown in a few studies \citep[see, e.g.,][]{2020MNRAS.492..405S,2021PhRvD.103j3023A}.

In {\tt relxill\_nk}, we have recently added the model {\tt relxillion\_nk}, which permits the user to model the radial ionization profile with a power-law \citep{2021PhRvD.103j3023A}. The model has an extra parameter, the ionization gradient index $\alpha_\xi$, and the ionization parameter is
\be
\xi(r)=\xi_{\rm in} \left( \frac{R_{\rm in}}{r} \right)^{\alpha_\xi},
\ee
where $\xi_{\rm in}$ is the value of the ionization parameter at the inner edge of the accretion disk $R_{\rm in}$. However, a similar model is not self-consistent because one infers the ionization and the emissivity profiles from the fit as two independent quantities, while $\xi$ and $\epsilon$ are connected by Eq.~(\ref{eq-xi}), where there is also the electron density $n$. The same criticism applies to the radial ionization profiles of the other reflection models used in the literature to date. In this paper, we thus present {\tt relxilldgrad\_nk}, which is a new flavor of the {\tt relxill\_nk} package to solve this issue and have a self-consistent model.

{\tt relxilldgrad\_nk} employs the grid of the non-relativistic reflection model {\tt xillverD} and the disk electron density profile is described by a power-law
\be\label{eq:N_profile}
n(r)=n_{\rm in} \left( \frac{R_{\rm in}}{r} \right)^{\alpha_n},
\ee
where $R_{\rm in}$ is still the radial coordinate of the inner edge of the accretion disk, $n_{\rm in}$ is the value of the electron density at $R_{\rm in}$, and $\alpha_n$ is the power-law index of the density gradient. Setting $ \alpha_n=0 $, we recover the standard constant density profile, while for $\alpha_n>0$ we have an accretion disk in which the electron density decreases as the radial coordinate $r$ increases.

The ionization parameter at any point of the disk, $\xi(r)$, is calculated self-consistently from the values of the electron density and the emissivity at that point of the disk as 
\be\label{eq:Xi_profile}
\xi(r) = \xi_{\rm max} \ \left[ \frac{4\pi \epsilon(r)}{n(r)} \right]_{\rm Norm},
\ee
where $\xi_{\rm max}$ and $\epsilon(r)$ are, respectively, the maximum value of the ionization parameter over the disk and the emissivity at the radial coordinate $r$. Note the expression in square brackets on the right hand side of Eq.~(\ref{eq:Xi_profile}) is normalized with respect to its maximum value reached on the accretion disk. Depending on $\alpha_n$, $n(r)$, and $\epsilon(r)$, the function $\xi(r)$ may not have the maximum at $R_{\rm in}$.

\begin{figure*}[t]
\begin{center}
    \includegraphics[width=17cm,trim={2cm 0cm 2cm 0cm},clip]{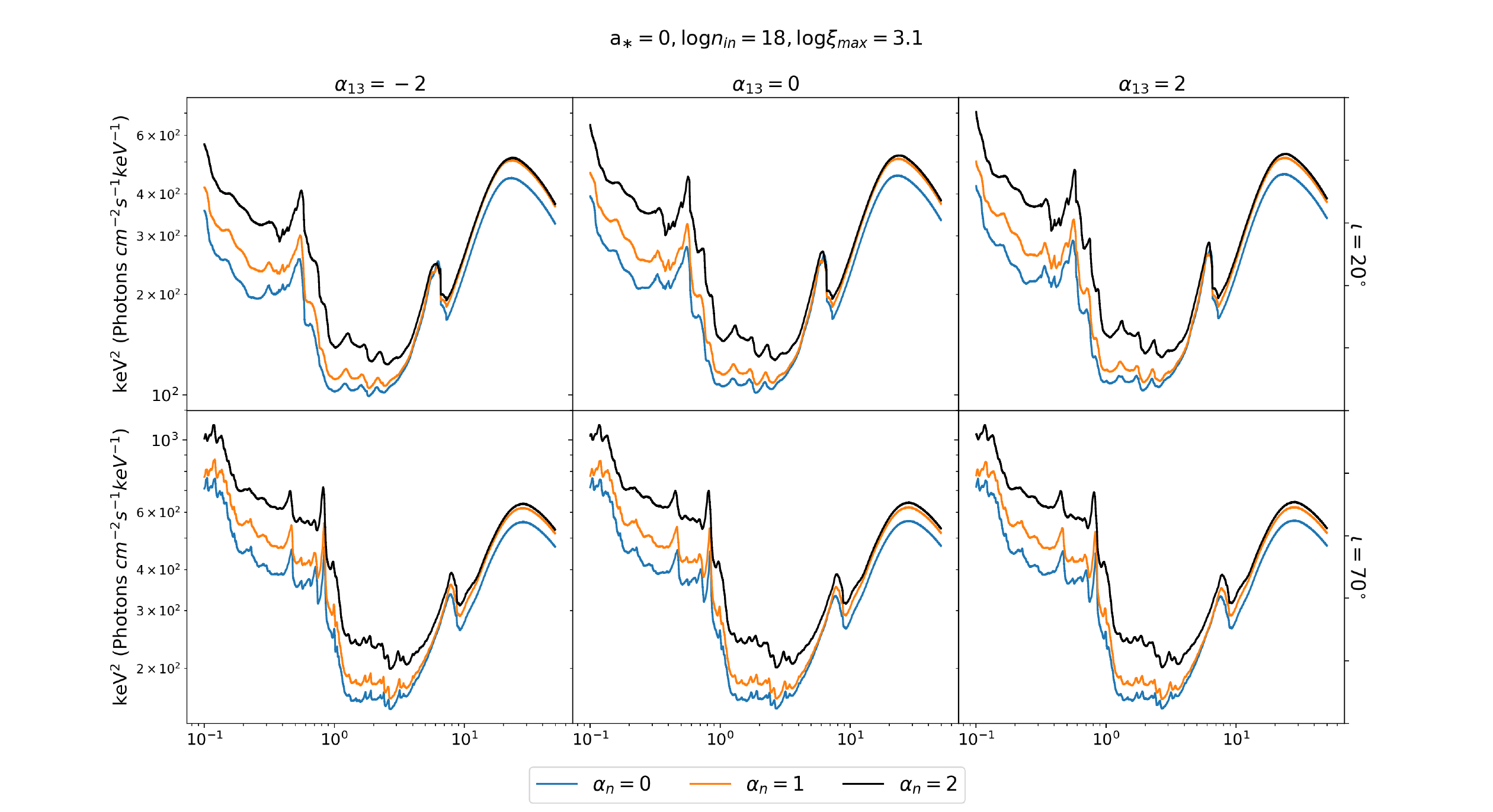}
\end{center}
\vspace{-0.2cm}
    \caption{Synthetic relativistic reflection spectra generated in the Johannsen spacetime for the spin parameter $a_* = 0$, electron density at the inner edge of the disk $\log n_{\rm in} = 18$, maximum value of the ionization parameter $\log \xi_{\rm max}=3.1$, deformation parameter $\alpha_{13}=-2$, 0, and 2, and inclination angle of the disk $\iota=20^\circ$ and $70^\circ$. For the emissivity profile, here we assume a power-law with emissivity index 3, i.e. $\epsilon \propto r^{-3}$. The spectra with constant electron density, $\alpha_n=0$, are in blue, those with density index $\alpha_n=1$ are in orange, and the spectra with $\alpha_n=2$ are in black. The photon index is $\Gamma = 2$ and we assume Solar iron abundance. \label{f-plot1}}
\vspace{0.5cm}
\begin{center}
    \includegraphics[width=17cm,trim={2cm 0cm 2cm 0cm},clip]{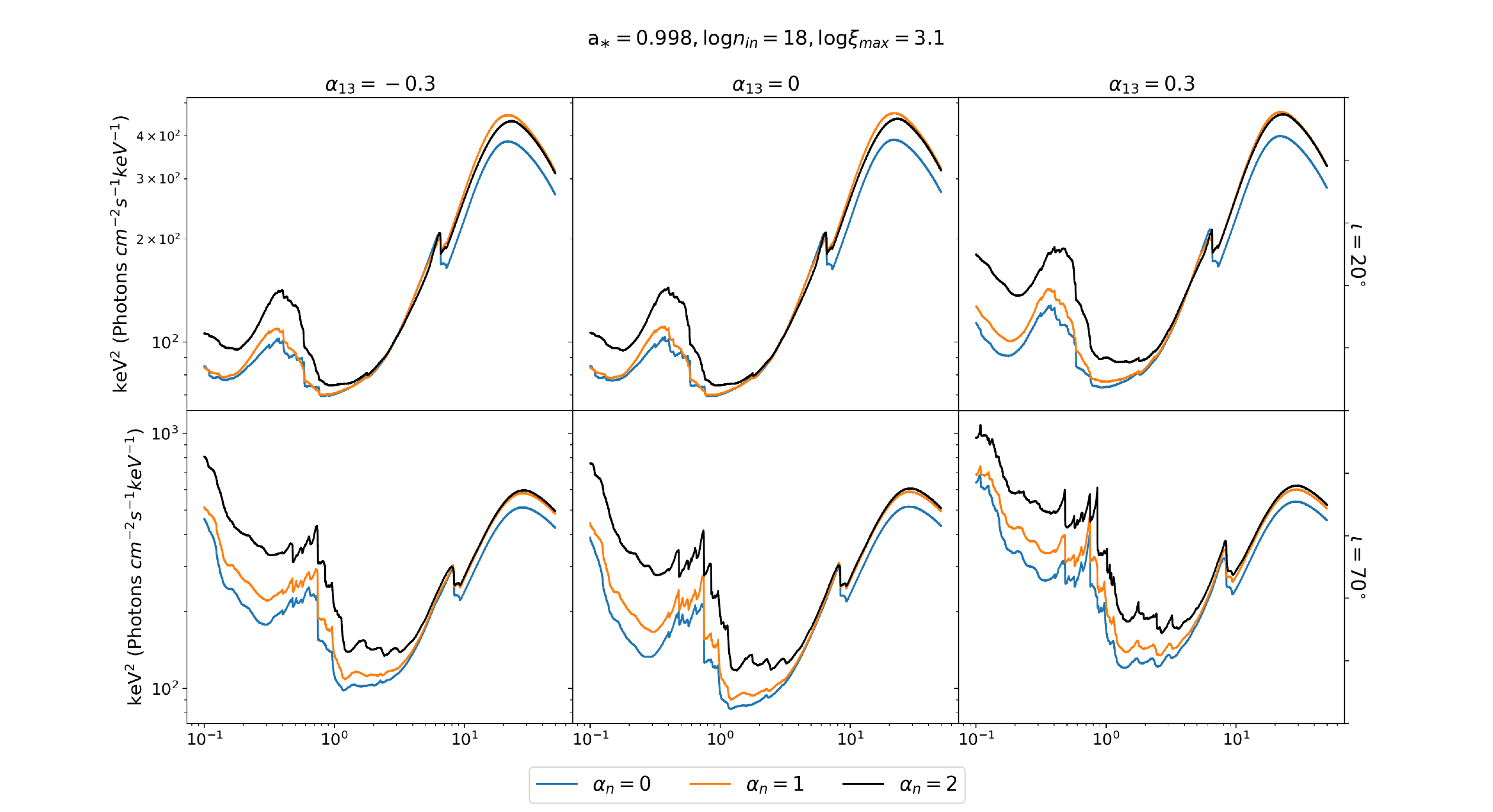}
\end{center}
\vspace{-0.2cm}
    \caption{As in Fig.~\ref{f-plot1} for the spin parameter $a_* = 0.998$ and the Johannsen deformation parameter $\alpha_{13}=-0.3$, 0, and 0.3. \label{f-plot2}}
\end{figure*}

\begin{figure*}[t]
\begin{center}
    \includegraphics[width=17cm,trim={2cm 0cm 2cm 0cm},clip]{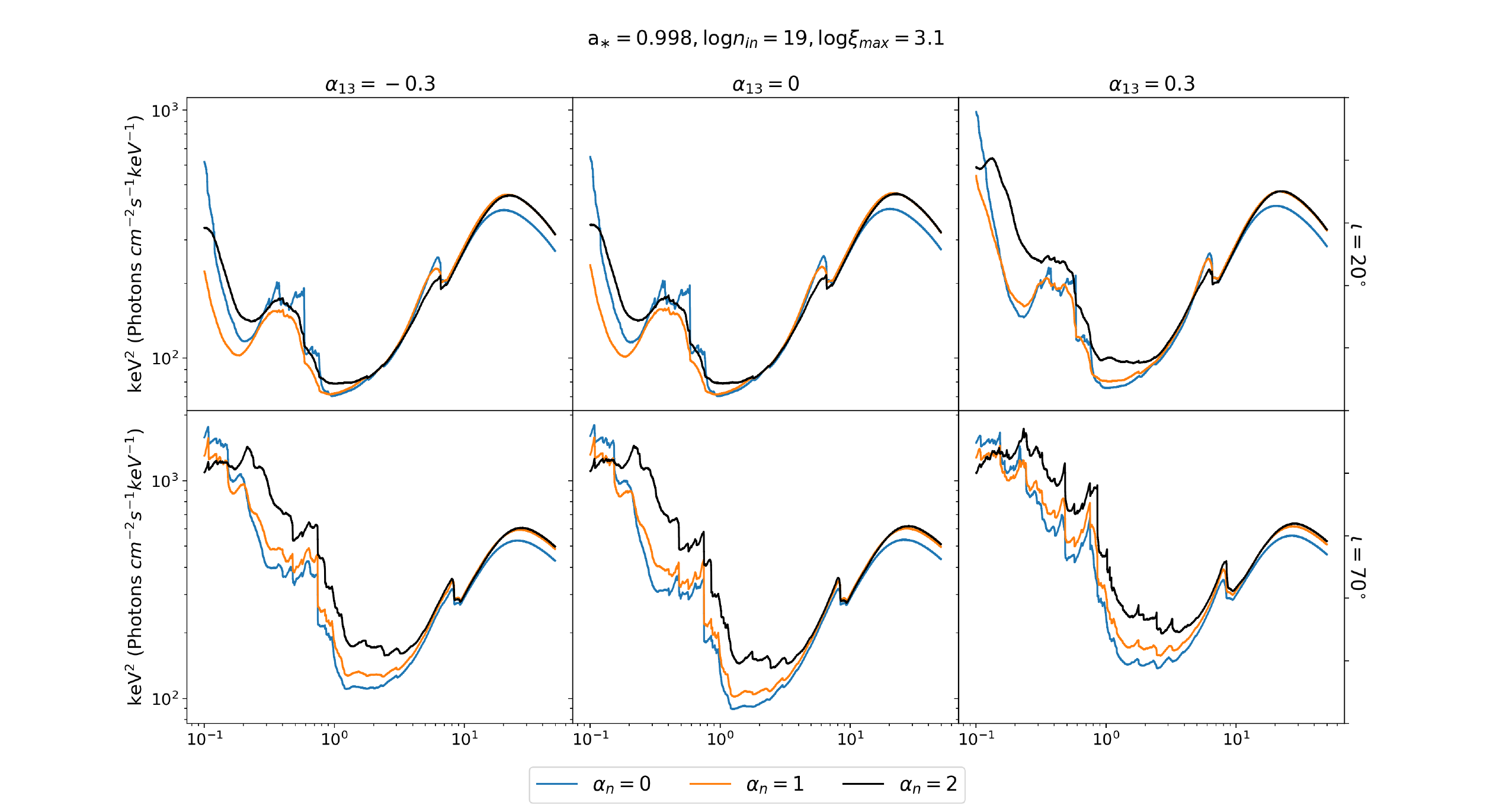}
\end{center}
\vspace{-0.2cm}
    \caption{As in Fig.~\ref{f-plot2} for $\log n_{\rm in} = 19$. \label{f-plot3}}
\vspace{0.5cm}
\begin{center}
    \includegraphics[width=17cm,trim={2cm 0cm 2cm 0cm},clip]{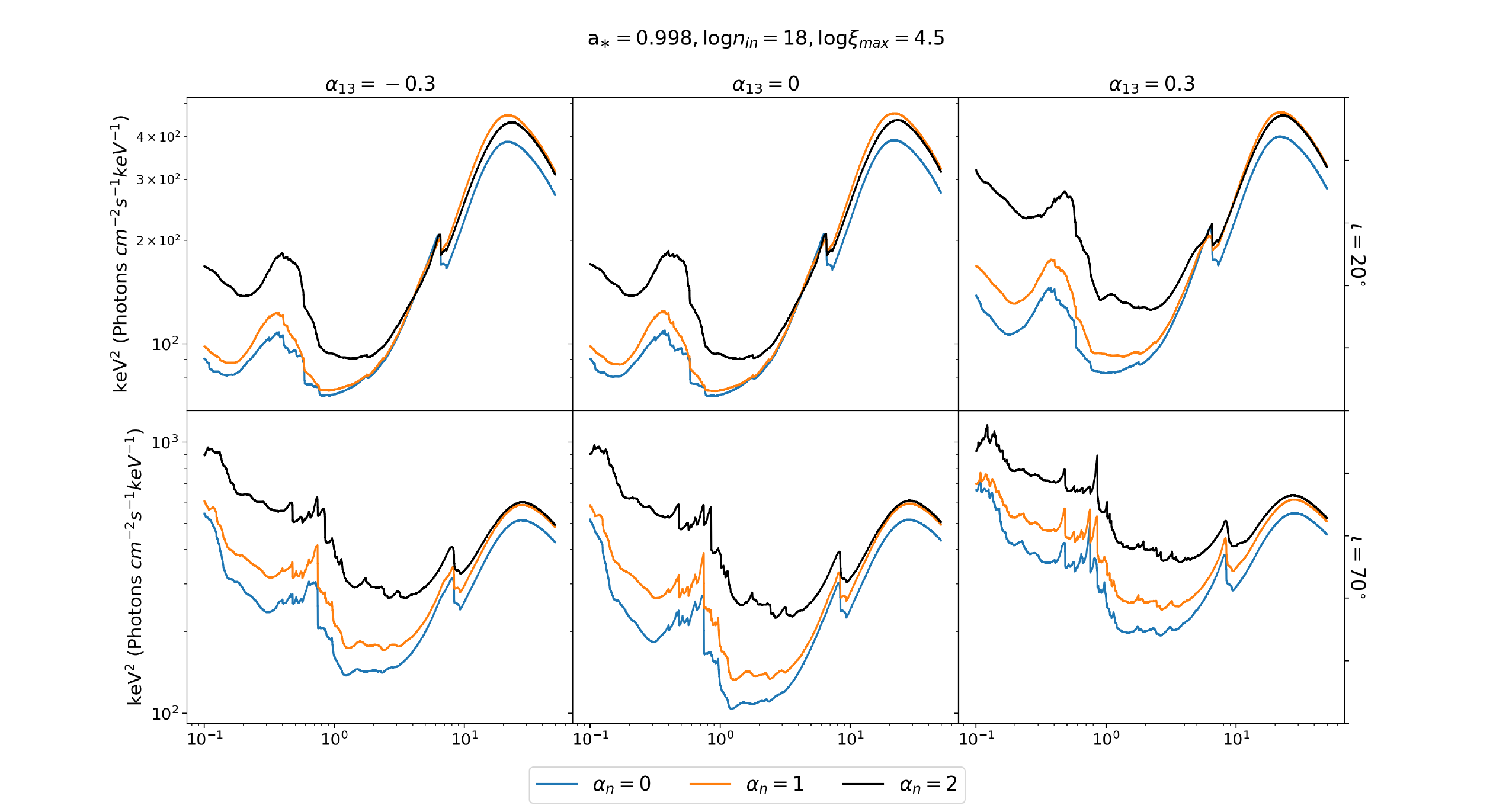}
\end{center}
\vspace{-0.2cm}
    \caption{As in Fig.~\ref{f-plot2} for $\log \xi_{\rm max} = 4.5$. \label{f-plot4}}
\end{figure*}

\begin{figure*}[t]
\begin{center}
    \includegraphics[width=17cm,trim={3cm 0cm 3cm 0cm},clip]{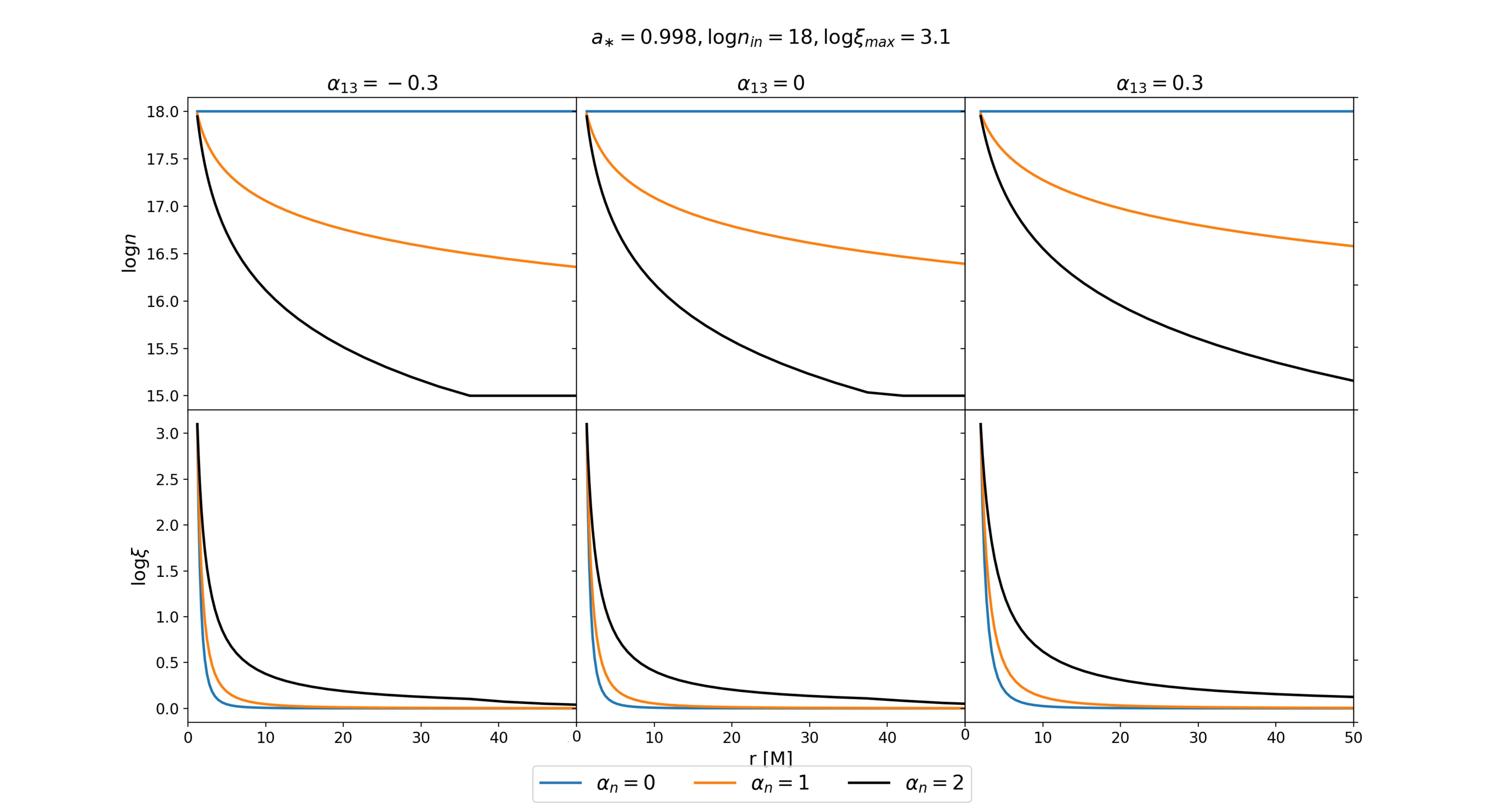}
\end{center}
\vspace{-0.2cm}
    \caption{Profiles of the electron density $n$ and of the ionization parameter $\xi$ for the astrophysical systems considered in Fig.~\ref{f-plot2}. \label{f-plot2nxi}}
    \vspace{0.5cm}
\end{figure*}

In the calculation process, the emissivity profile for the entire disk is calculated first. Then, we divide the accretion disk into 50 annuli and the values of the electron density and of the ionization parameter are calculated in the middle of each annulus using the already calculated emissivity profile values. Then, in each annulus, the reflection spectrum is extracted from the {\tt xillverD} table according to the values of the density and ionization parameters of this annulus. Finally, the reflection components from each annulus are convolved using the transfer functions corresponding to that annulus to obtain the reflection spectrum far away from the source, and the total spectrum is obtained by summing all the convolved spectra together.

Figs.~\ref{f-plot1}, \ref{f-plot2}, \ref{f-plot3}, and \ref{f-plot4} illustrate the impact of an electron density gradient on the reflection spectrum of a source. In all plots the inner edge of the accretion disk is assumed to be at the innermost stable circular orbit (ISCO), so $R_{\rm in} = R_{\rm ISCO}$ in Eq.~(\ref{eq:N_profile}). The constant electron density $\alpha_n=0$ (blue curves) is compared with the spectra calculated for the density index $\alpha_n=1$ (orange curves) and $\alpha_n=2$ (black curves) for various values of the black hole spin parameter $a_*$, the deformation parameter $\alpha_{13}$ of the Johannsen spacetime \citep[which is the default metric in {\tt relxill\_nk} and was proposed in][specifically for testing the Kerr metric with black hole electromagnetic data]{2013PhRvD..88d4002J}, the inclination angle $\iota$, the electron density at the inner edge of the disk $n_{\rm in}$, and the maximum value of the ionization parameter $\xi_{\rm max}$. The value of $\alpha_n$ has a larger impact on the shape of the spectrum below the iron line because that is the region of the fluorescent emission lines, and the energy and the strength of these lines are quite sensitive to the ionization parameter. Fig.~\ref{f-plot2nxi} shows the profiles of the electron density and of the ionization parameter for the systems in Fig.~\ref{f-plot2} ($a_* = 0.998$, $\log n_{\rm in} = 18$, and $\log\xi_{\rm max} = 3.1$).


\section{Implications on spectral fitting} \label{s-app}

In this section we show the impact of a radial electron density profile on spectral fitting. First, we fit a \textsl{NuSTAR} spectrum of the Galactic black hole EXO~1846--031. Then, we simulate a simultaneous observation of the same source with \textsl{Athena} and \textsl{eXTP}.

\subsection{\textsl{NuSTAR} spectrum of EXO~1846--031}

EXO~1846--031 was discovered by \textsl{EXOSAT} on 1985 April 3~\citep{1985IAUC.4051....1P} and was later identified as a low mass X-ray binary harboring a black hole candidate~\citep{1993A&A...279..179P}. A second outburst was observed in 1994~\citep{1994IAUC.6096....1Z}. A third and, to date, last outburst was first detected with \textsl{MAXI} in July 2019~\citep{2019ATel12968....1N} and the source was then observed with several other instruments. \textsl{NuSTAR} observed EXO~1846--031 on 2019 August 3 (Observation ID 90501334002) and in what follows we analyze such data. \citet{2020ApJ...900...78D} were the first to analyze this observation and here we will follow their analysis. We note that the source was very bright during this observation and that previous studies found that the black hole has a spin parameter very close to 1 and that the viewing angle of the accretion disk is very high, two ingredients that maximize relativistic effects. This \textsl{NuSTAR} spectrum is quite simple, with a very broadened and prominent iron line. All these properties make this \textsl{NuSTAR} observation of EXO~1846--031 particularly suitable for testing fundamental physics from the analysis of the reflection features as well as for testing new reflection models. This observation was analyzed with {\tt relxill\_nk} to test the Kerr metric in \citet{2021ApJ...913...79T}.

For the analysis, we use XSPEC v12.11.1~\citep{1996ASPC..101...17A}, WILMS abundances~\citep{2000ApJ...542..914W}, and VERNER cross-sections~\citep{1996ApJ...465..487V}. The \textsl{NuSTAR} observation is 22~ks. The flux of the source does not show significant variability, and we can thus use the time-averaged spectrum for the spectral analysis. The unprocessed raw data are downloaded from the HEASARC website. HEASOFT module {\tt nupipeline} is used to process the 
unfiltered data into cleaned event data. A circular region of 180~arcsec is taken around the center
of the source to extract the spectrum. A background region of the same size is taken as far as possible from the source on the same detector. For the extraction of source and background spectra, another HEASOFT module, {\tt nuproducts}, is used. The same module is used to generate the response matrix file and the ancillary file. We rebin the source spectra to have at least 30 counts per bin in order to use $\chi^2$ statistics.

\begin{figure}[t]
\begin{center}
    \includegraphics[width=8.5cm,trim={2cm 0.5cm 4cm 18cm},clip]{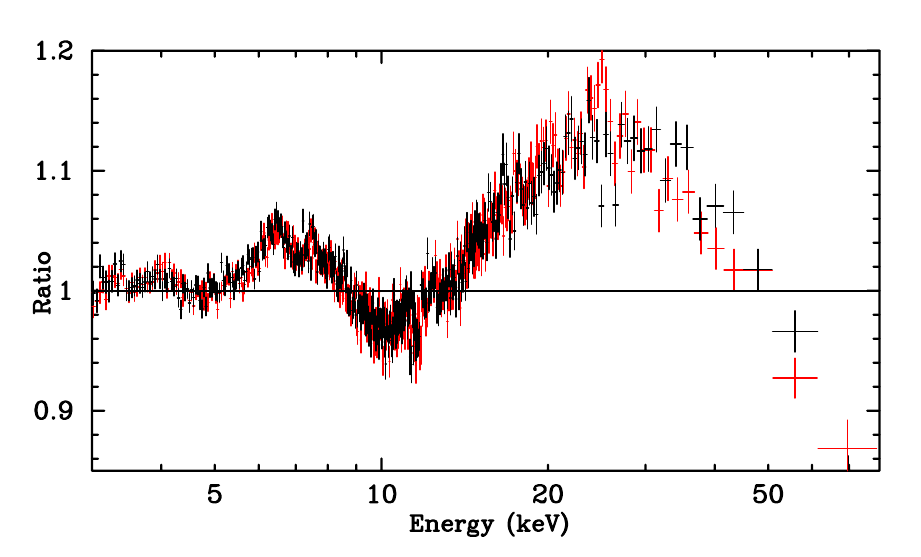}
\end{center}
\vspace{-0.5cm}
    \caption{Data to best-fit model ratio for an absorbed power-law; in XSPEC language, {\tt tbabs$\times$powerlaw}. Red crosses are for FPMA data and black crosses for FPMB data. \label{f-ref}}
    \vspace{0.5cm}
\end{figure}

If we fit the data with an absorbed power-law, {\tt tbabs$\times$powerlaw} in XSPEC language, we find the data to best-fit model ratio shown in Fig.~\ref{f-ref}. We clearly see a prominent and broadened iron line in the soft X-ray band and a Compton hump peaked at 20-30~keV, indicating that the spectrum has strong reflection features.

We fit the data with three reflection models: normal {\tt relxill\_nk} with constant ionization parameter and electron density over the whole disk, {\tt relxillion\_nk} with a radial ionization profile and a constant electron density, and {\tt relxilldgrad\_nk} with a radial electron density profile and a self-consistent ionization profile calculated from Eq.~(\ref{eq:Xi_profile}). We note the key-differences among the three models. {\tt relxill\_nk} and {\tt relxillion\_nk} use the {\tt xillver} table, so the electron density is fixed to $10^{15}$~cm$^{-3}$ and the high-energy cutoff in the coronal spectrum $E_{\rm cut}$ is free. When the ionization gradient index $\alpha_\xi$ in {\tt relxillion\_nk} is set to zero, {\tt relxillion\_nk} reduces exactly to {\tt relxill\_nk}. On the other hand, {\tt relxilldgrad\_nk} uses the table of {\tt xillverD}, where the electron density is free and the high-energy cutoff $E_{\rm cut}$ is fixed to 300~keV. So {\tt relxilldgrad\_nk} reduces exactly to {\tt relxill\_nk} only when its density gradient index $\alpha_n$ is set to zero and, at the same time, we set $E_{\rm cut} = 300$~keV in {\tt relxill\_nk}. In the three fits, we model the emissivity profile with a broken power-law (three free parameters: the inner emissivity index $q_{\rm in}$, the outer emissivity index $q_{\rm out}$, and the breaking radius $R_{\rm br}$) and we impose that the inner edge of the disk $R_{\rm in}$ is at the ISCO radius. The black hole spin parameter $a_*$, the inclination angle of the disk $i$, the photon index $\Gamma$ and the high-energy cutoff $E_{\rm cut}$ of the coronal spectrum, the iron abundance $A_{\rm Fe}$, and the deformation parameter of the spacetime $\alpha_{13}$ are always free in the fits.

When we use {\tt relxill\_nk}, the XSPEC model is

\vspace{0.2cm}

{\tt tbabs$\times$(diskbb + relxill\_nk + gaussian)} .

\vspace{0.2cm}

\noindent This is the same fit as in \citet{2020ApJ...900...78D}, with the difference that \citet{2020ApJ...900...78D} used {\tt relxill} that assumes the Kerr metric and here we use {\tt relxill\_nk} with a free deformation parameter $\alpha_{13}$. {\tt diskbb} is introduced to fit a weak thermal component of the accretion disk~\citep{1984PASJ...36..741M} and the temperature of the inner edge of the disk and the normalization are free in the fit.
{\tt gaussian} is necessary to fit an unresolved feature (but it only improves the fit, without any significant impact on the estimate of the model parameters). 
The width of the Gaussian and the reflection fraction $R_{\rm f}$ in {\tt relxill\_nk} cannot be constrained simultaneously. In our fit, we allow the width of the Gaussian to vary in the range 1-100~eV (the best-fit is around 70~eV but the parameter is still unconstrained) and the reflection fraction is frozen to 1.

\begin{figure}
\begin{center}
\includegraphics[width=8cm,trim={1.5cm 2.5cm 1cm 1.0cm},clip]{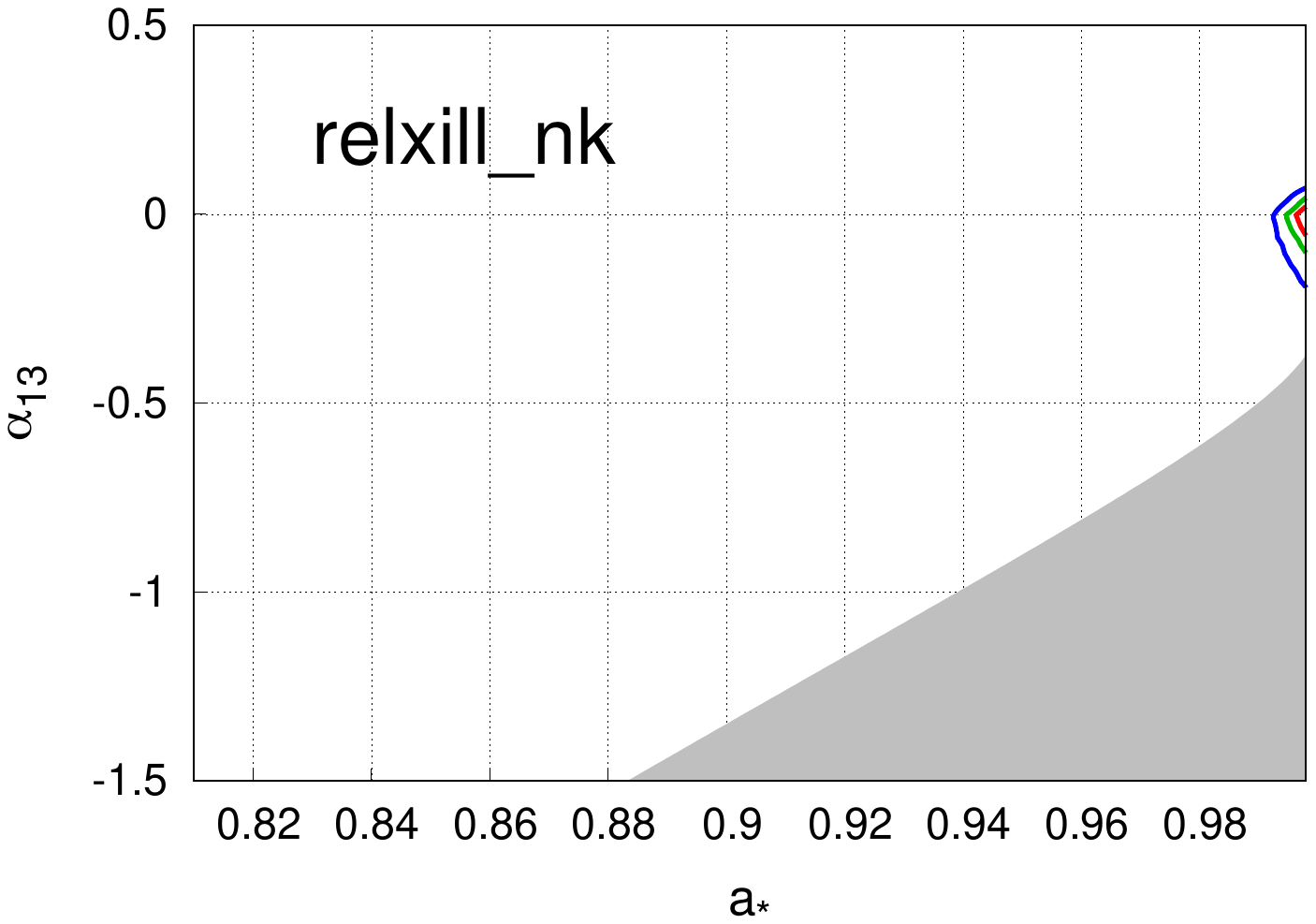}\\
\includegraphics[width=8cm,trim={1.5cm 2.5cm 1cm 0.5cm},clip]{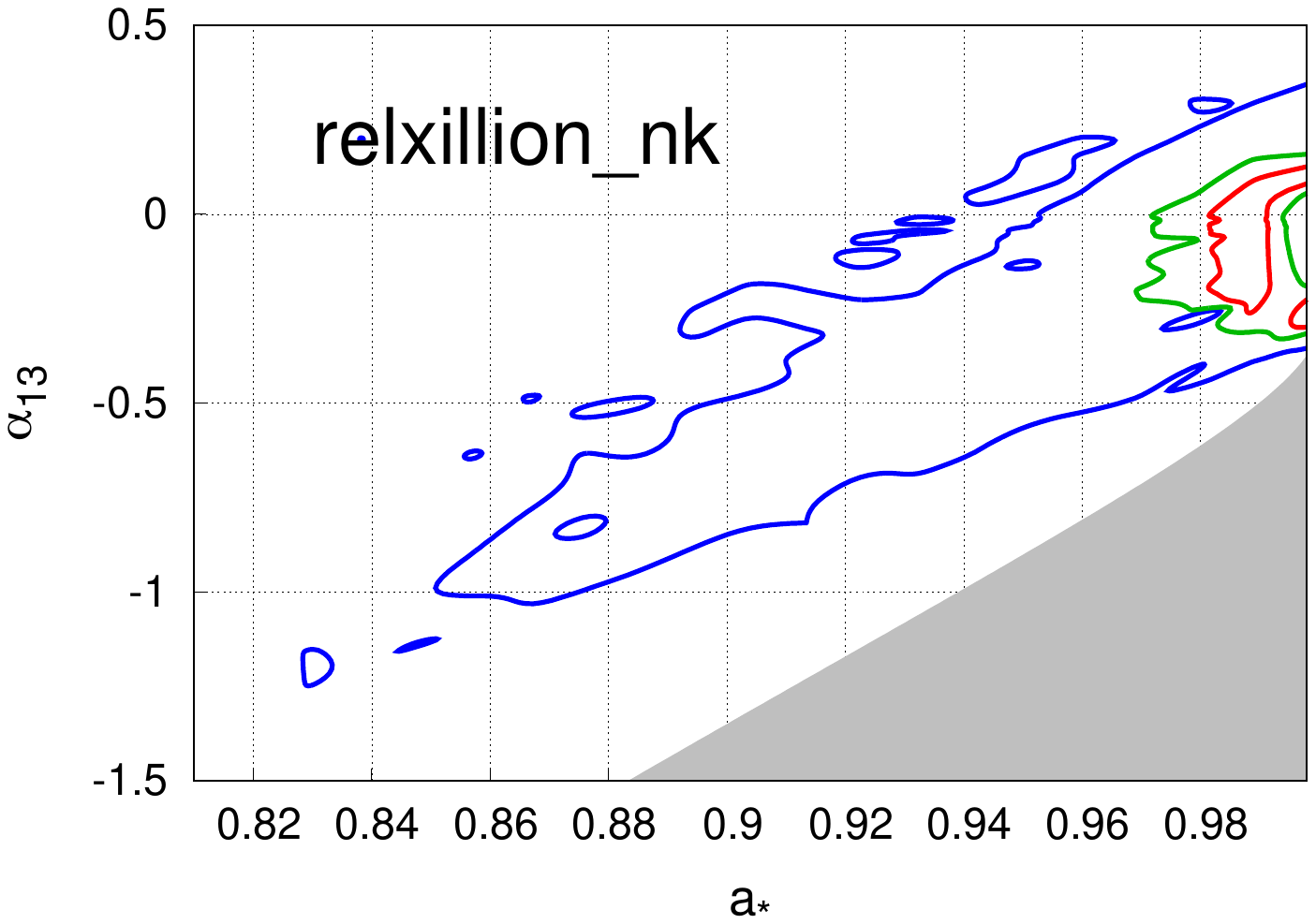}\\
\includegraphics[width=8cm,trim={1.5cm 2.5cm 1cm 0.5cm},clip]{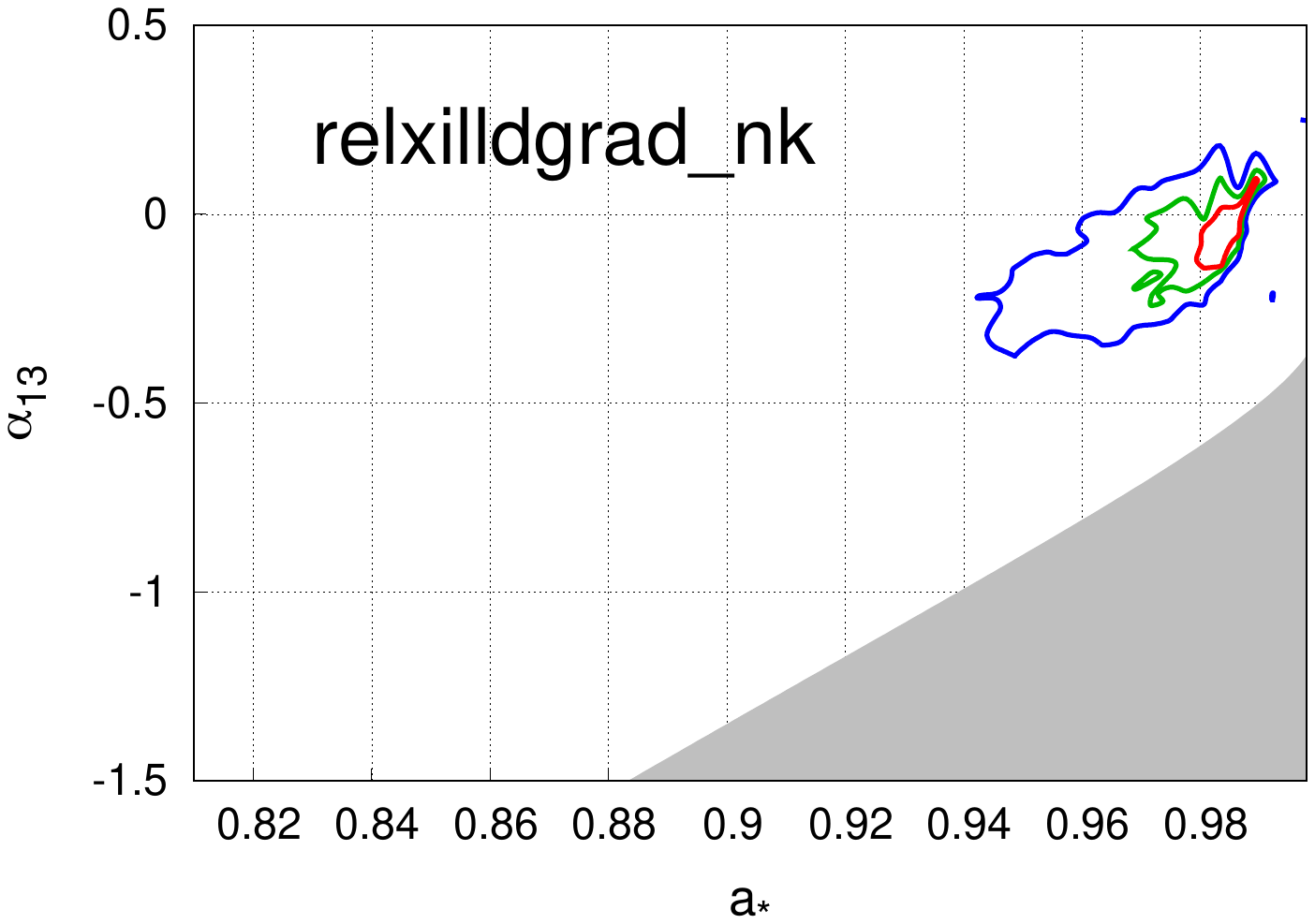}
\end{center}
\vspace{-0.2cm}
\caption{Constraints on the black hole spin parameter $a_*$ and on the deformation parameter $\alpha_{13}$ obtained with {\tt relxill\_nk}, {\tt relxillion\_nk}, and {\tt relxilldgrad\_nk} from the analysis of the 2019 \textsl{NuSTAR} data of EXO~1846--031. The red, green, and blue curves correspond, respectively, to the 68\%, 90\%, and 99\% confidence levels for two relevant parameters. The gray region is not included in our analysis because the spacetime is not regular there \citep[see the discussion in][for more details]{2020arXiv201207469R}.  \label{f-exo}}
\end{figure}

Interestingly, when we use {\tt relxillion\_nk} and {\tt relxilldgrad\_nk}, we do not need {\tt gaussian} and we can leave the reflection fraction free, as the fit itself requires a reflection fraction close to 1. Since {\tt relxilldgrad\_nk} uses the table of {\tt xillverD}, which is for a variable electron density $n$ but assumes a constant high-energy cutoff in the coronal spectrum ($E_{\rm cut} = 300$~keV) in order to limit the size of the table, when we use {\tt relxilldgrad\_nk} we freeze its reflection fraction to $-1$ (no coronal spectrum in the output) and we add {\tt cutoffpl} to describe the continuum and have a variable high-energy cutoff. In the end, the two XSPEC models are, respectively,

\vspace{0.2cm}

{\tt tbabs$\times$(diskbb + relxillion\_nk}) ,

\vspace{0.2cm}

{\tt tbabs$\times$(diskbb + relxilldgrad\_nk + cutoffpl)} ,

\vspace{0.2cm}

The best-fit values of the three cases are reported in Tab.~\ref{t-nustar}, left side. The data to best-fit model ratios are shown in the left panel of Fig.~\ref{ratio}. Fig.~\ref{f-exo} shows the constraints on the black hole spin parameter $a_*$ and on the deformation parameter $\alpha_{13}$ inferred from the three fits. We postpone the discussion of these results to Section~\ref{s-dc}.

\subsection{Simulation with \textsl{Athena} and \textsl{eXTP}}

\begin{figure*}
\begin{center}
\includegraphics[width=8cm]{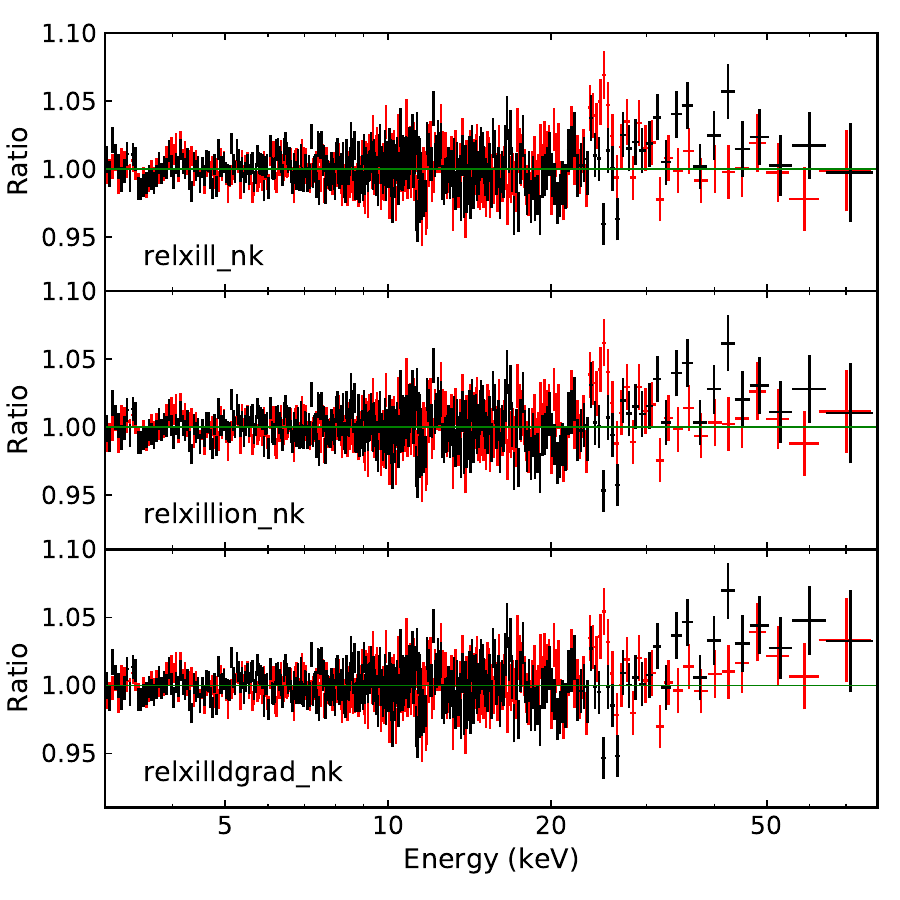}
\hspace{0.8cm}
\includegraphics[width=8cm]{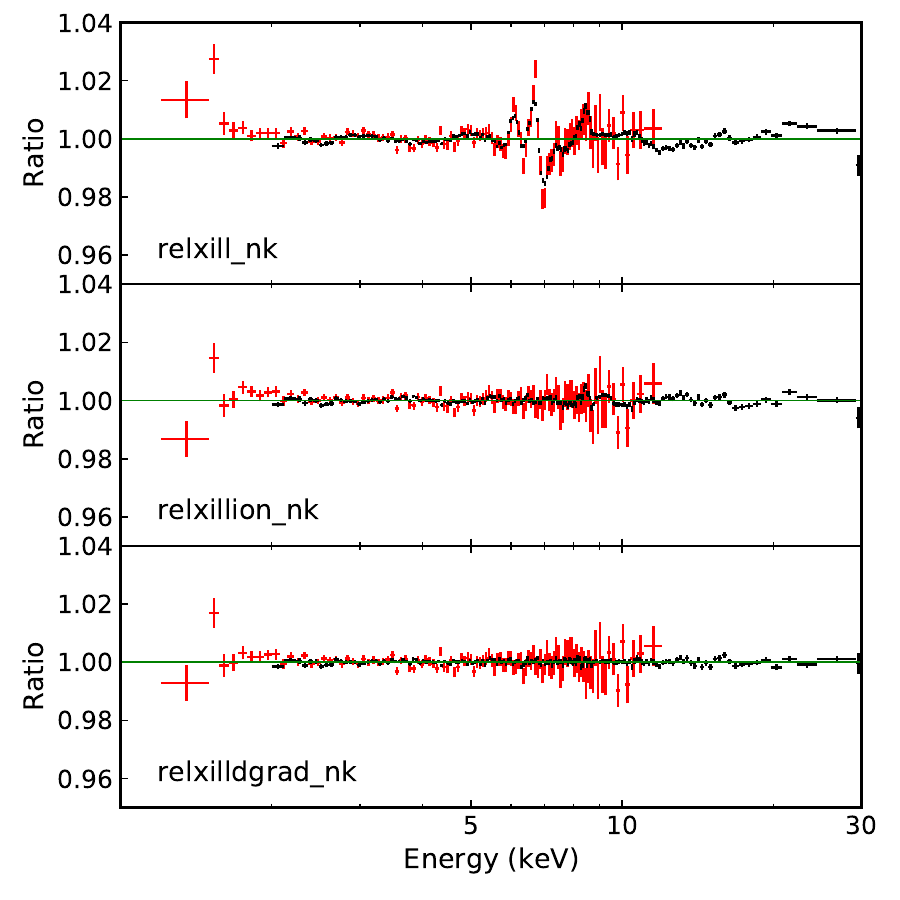}
\end{center}
\vspace{-0.3cm}
\caption{Data to model ratio for the analysis of \textsl{NuSTAR} data (left panel; black crosses for FPMA and red crosses for FPMB) and the simulated observation with \textsl{Athena}+\textsl{eXTP} (right panel; red crosses for \textsl{Athena} and black crosses for \textsl{eXTP}) of EXO~1846--031.}
\label{ratio}
\end{figure*}

In order to see the differences of the three models in the previous subsection for the next generation of X-ray missions, we simulate a simultaneous observation with \textsl{Athena}~\citep{2013arXiv1306.2307N} and \textsl{eXTP}~\citep{2016SPIE.9905E..1QZ} of EXO~1846--031. We assume a 20~ks observation with the X-ray Integral Field Unit (X-IFU) for \textsl{Athena} and with the Large Area Detector (LAD) for \textsl{eXTP}. The best-fit values of the {\tt relxilldgrad\_nk} model found in the previous subsection are used as input for the simulation, with the exception of the deformation parameter which is assumed to vanish: $\alpha_{13} = 0$ (Kerr metric). With such a prescription, we obtain about $3\times 10^7$ and $3\times 10^8$ counts for X-IFU and LAD, respectively. We fit the simulated spectra with the reflection models {\tt relxill\_nk}, {\tt relxillion\_nk}, and {\tt relxilldgrad\_nk}. We note that now, when we use {\tt relxill\_nk}, we do not need {\tt gaussian} and we can leave the reflection fraction free, as for the other two fits. The best-fit parameters and residuals for the simulation are shown, respectively, in Tab.~\ref{t-nustar}, right side, and Fig.~\ref{ratio}, right panel. We do not show here the confidence level curves on the plane black hole spin parameter $a_*$ vs. deformation parameter $\alpha_{13}$ because the 68\%, 90\%, and 99\% confidence regions are too small, but the constraints on $a_*$ and $\alpha_{13}$ can be seen from Tab.~\ref{t-nustar}. The discussion of these results is presented in the next section.

\begin{table*}
\centering
{\renewcommand{\arraystretch}{1.3}
\begin{tabular}{lccc|ccc}
\hline\hline
&& \textsl{NuSTAR} &&& \textsl{Athena}+\textsl{eXTP} & \\
Model & {\tt relxill\_nk} & {\tt relxillion\_nk} & {\tt relxilldgrad\_nk} & {\tt relxill\_nk} & {\tt relxillion\_nk} & {\tt relxilldgrad\_nk} \\
\hline\hline
{\tt tbabs} \\
$N_{\rm H}$ [$10^{22}$~cm$^{-2}$] & $10.8_{-0.5}^{+0.7}$ & $7.7_{-0.9}^{+0.8}$ & $8.4_{-0.4}^{+0.6}$ & $8.547_{-0.007}^{+0.003}$ & $8.422_{-0.009}^{+0.010}$ & $8.456_{-0.007}^{+0.008}$ \\
\hline 
{\tt diskbb} \\
$kT_{\rm in}$ [keV] & $0.434_{-0.010}^{+0.009}$ & $0.430_{-0.029}^{+0.008}$ & $0.387_{-0.027}^{+0.021}$ & $0.37806_{-0.00020}^{+0.00040}$ & $0.3929_{-0.0012}^{+0.0007}$ & $0.3862_{-0.0007}^{+0.0004}$ \\
\hline
{\tt relxill\_nk} \\
$q_{\rm in}$ & $8.58_{-0.55}^{+0.20}$ & $9.9_{-0.8}^{\rm + (P)}$ & $7.98_{-0.23}^{+0.09}$ & $10_{-0.022}$ & $6.42_{-0.28}^{+0.22}$ & $7.9663_{-0.0012}^{+0.0379}$ \\
$q_{\rm out}$ & $0.1_{\rm -(P)}^{+3.3}$ & $0_{}^{+0.20}$ & $0_{}^{+0.14}$ & $0^{+0.003}$ & $0.49_{-0.04}^{+0.09}$ & $0^{+0.03}$ \\
$R_{\rm br}$ [$r_{\rm g}$] & $6.1_{-2.8}^{+0.5}$ & $7.4_{-2.8}^{+2.8}$ & $8.3_{-0.5}^{+0.3}$ & $6.521_{-0.031}^{+0.010}$ & $8.8_{-0.5}^{+0.5}$ & $8.28_{-0.10}^{+0.07}$ \\
$a_*$ & $0.998_{-0.0015}^{}$ & $0.988_{-0.010}^{+0.005}$ & $0.9852_{-0.0056}^{+0.0020}$ & $0.998_{-0.0009}$ & $0.9978_{-0.0012}^{\rm +(P)}$ & $0.9862_{-0.0003}^{+0.0003}$ \\
$i$ [deg] & $76.7_{-0.4}^{+1.1}$ & $68.4_{-0.4}^{+0.4}$ & $73.53_{-0.19}^{+1.02}$ & $71.419_{-0.116}^{+0.019}$ & $71.50_{-0.32}^{+0.23}$ & $73.70_{-0.17}^{+0.23}$ \\
$\Gamma$ & $1.992_{-0.004}^{+0.006}$ & $1.781_{-0.015}^{+0.055}$ & $1.984_{-0.022}^{+0.025}$ & $2.045_{-0.003}^{+0.005}$ & $1.949_{-0.006}^{+0.008}$ & $1.991_{-0.003}^{+0.005}$ \\
$E_{\rm cut}$ [keV] & $200_{-6}^{+5}$ & $143_{-6}^{+19}$ & $58_{-15}^{+7}$ & $241.8_{-3.0}^{+1.7}$ & $191_{-3}^{+4}$ & $65_{-3}^{+4}$ \\
$\log\xi_{\rm max}$ [erg~cm~s$^{-1}$] & $3.518_{-0.047}^{+0.023}$ & $4.7_{-0.05}^{}$ & $3.96_{-0.16}^{+0.32}$ & $3.298_{-0.003}^{+0.004}$ & $4.293_{-0.054}^{+0.023}$ & $3.948_{-0.045}^{+0.022}$ \\
$\alpha_\xi$ & $0^\star$ & $1.82_{-0.16}^{+0.28}$ & -- & $0^\star$ & $1.09_{-0.03}^{+0.03}$ & -- \\
$\log n_{\rm in}$ [cm$^{-3}$] & $15^\star$ & $15^\star$ & $16.32_{-0.04}^{+0.03}$ & $15^\star$ & $15^\star$ & $16.323_{-0.006}^{+0.004}$ \\
$\alpha_n$ & $0^\star$ & $0^\star$ & $8.30_{-0.24}^{+0.21}$ & $0^\star$ & $0^\star$ & $8.31_{-0.05}^{+0.04}$ \\
$A_{\rm Fe}$ & $0.835_{-0.033}^{+0.014}$ & $8.8_{-2.2}^{+0.4}$ & $1.28_{-0.16}^{+0.21}$ & $1.003_{-0.004}^{+0.006}$ & $1.89_{-0.15}^{+0.16}$ & $1.306_{-0.031}^{+0.023}$ \\
$R_{\rm f}$ & $1^\star$ & $1.09_{-0.18}^{+0.26}$ & -- & $0.1583_{-0.0012}^{+0.0011}$ & $1.21_{-0.13}^{+0.21}$ & -- \\
$\alpha_{13}$ & $0.00_{-0.06}^{+0.03}$ & $-0.01_{-0.26}^{+0.15}$ & $-0.03_{-0.12}^{+0.13}$ & $0.1576_{-0.0011}^{+0.007}$ & $-0.245_{-0.009}^{+0.014}$ & $0.00_{-0.04}^{+0.08}$ \\
\hline
{\tt gaussian} \\
$E_{\rm line}$ & $7.02_{-0.03}^{+0.05}$ & -- & -- & -- & -- & -- \\
\hline
$\chi^2/{\rm dof}$ & 2754.13/2594 & 2715.32/2595 & 2722.12/2594 & 27679.74/ 22845 & 23454.58/22844 & 23248.13/22843 \\
& =1.06173 & =1.04637 & =1.04939 & =1.211632 & =1.026728 & =1.017736 \\ 
\hline\hline
\end{tabular}
}
\caption{\rm Summary of the analysis of the 2019 \textsl{NuSTAR} observation of EXO~1846--031 and of the \textsl{Athena}+\textsl{eXTP} simulation. The reported uncertainties correspond to the 90\% confidence level for one relevant parameter ($\Delta\chi^2 = 2.71$). $^\star$ indicates that the value of the parameter is frozen in the fit. (P) means that the 90\% confidence level reaches the upper/lower boundary of the parameter. If there is no upper/lower uncertainty, the best-fit value is stuck at the upper/lower boundary of the parameter. $q_{\rm in}$ and $q_{\rm out}$ are allowed to vary in the range 0 to 10. The maximum value of $a_*$ is 0.998. The maximum value of $\log\xi_{\rm max}$ is 4.7. \label{t-nustar}}
\end{table*}

\section{Discussion and conclusions} \label{s-dc}

In this paper, we have presented {\tt relxilldgrad\_nk}, which is an extension of the {\tt relxill\_nk} package in which the disk electron density profile is described by a power-law and the ionization parameter is calculated self-consistently from the electron density and the emissivity. In order to see the impact of this implementation on the fits of reflection dominated spectra, in the previous section we have used {\tt relxill\_nk}, {\tt relxillion\_nk}, and {\tt relxilldgrad\_nk} to fit a \textsl{NuSTAR} observation of EXO~1846--031 (as an example of current X-ray observation) and the simulated simultaneous observation \textsl{Athena}+\textsl{eXTP} of the same source (as the prototype of observation possible with the next generation of X-ray missions).

Concerning the fits of the \textsl{NuSTAR} spectrum, from the ratio plots (left panel in Fig.~\ref{ratio}), we do not see clear differences among the three fits. If we consider the value of $\chi^2$, we see that the $\chi^2$ of the fit with {\tt relxill\_nk} is a bit higher than the other two. However, when we use {\tt relxill\_nk} we need to add a Gaussian and to freeze the reflection fraction to 1, while with the other two models these issues are not present: the Gaussian is not necessary and the data require a reflection fraction close to 1 (in the fit with {\tt relxilldgrad\_nk}, we do not have the reflection fraction, but we can estimate this quantity from the fluxes in the reflection and power-law components). In the end, we can argue that {\tt relxillion\_nk} and {\tt relxilldgrad\_nk} provide a better description of these data. If we compare {\tt relxillion\_nk} and {\tt relxilldgrad\_nk}, we see that the $\chi^2$ of {\tt relxillion\_nk} is slightly lower than that of {\tt relxilldgrad\_nk}. On the other hand, the fit with {\tt relxillion\_nk} recovers an unnaturally high iron abundance, suggesting some problem in the fit. We also note a discrepancy in the estimate of $E_{\rm cut}$ between the two models. It is remarkable that the estimate of the black hole spin parameter, $a_*$, and of the Johannsen deformation parameter, $\alpha_{13}$, is consistent among the three models. In other words, the measurement of these two parameters does not seem to be affected by the profiles of the ionization parameter and of the electron density, at least for these high-quality \textsl{NuSTAR} data.

\begin{figure}
\begin{center}
\includegraphics[width=8cm]{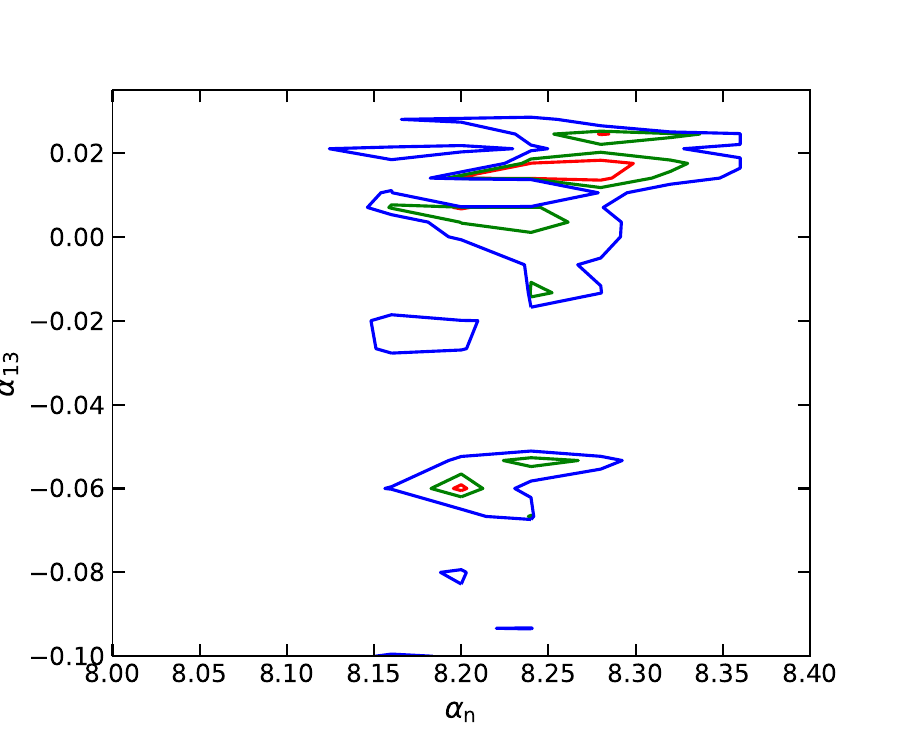}
\end{center}
\vspace{-0.5cm}
\caption{Study of the possible correlation between the deformation parameter $\alpha_{13}$ and the gradient density index $\alpha_n$ from the simulated observation \textsl{Athena}+\textsl{eXTP} when we fit the data with {\tt relxilldgrad\_nk}. The red, green, and blue curve represent, respectively, the 68\%, 90\%, and 99\% confidence level limits for two relevant parameters.}
\label{a13an}
\end{figure}

From the simulated observation \textsl{Athena}+\textsl{eXTP}, we can learn something more. First, from the ratio plots in the right panel of Fig.~\ref{ratio} we clearly see that {\tt relxill\_nk} is unsuitable to fit the data and there are important residuals in the iron line region. On the contrary, {\tt relxillion\_nk} can fit the data well. These conclusions are confirmed by the values of the reduced $\chi^2$. {\tt relxill\_nk} cannot fit these simulated data well, while both {\tt relxillion\_nk} and {\tt relxilldgrad\_nk} can (but in the case of {\tt relxilldgrad\_nk} it was expected because the simulation uses {\tt relxilldgrad\_nk}). This is quite an important point: even the high-quality data of \textsl{Athena}+\textsl{eXTP} may not be able to discriminate the two models from the quality of the fits\footnote{Such a statement is based on the spectral analysis of our study. However, \textsl{Athena} and especially \textsl{eXTP} data will be excellent for timing analyses and modeling timing properties may well be able to break this degeneracy. For instance, \citet{2019MNRAS.488..324I} show that models with a constant ionization parameter and with an ionization gradient can have similar spectra but they can be distinguished from their different variability properties.}.
We still see the difference in the estimate of $E_{\rm cut}$, as in the case of the \textsl{NuSTAR} spectrum. Last but not least, we see that the estimates of the black hole spin parameter are not very different among the three fits (so reliable spin measurements may still be obtained with a reflection model with constant ionization parameter and electron density), while we find that neither {\tt relxill\_nk} nor {\tt relxillion\_nk} can get an accurate measurement of the deformation parameter $\alpha_{13}$. This is the most important finding of the present work. Since {\tt relxillion\_nk} provides a good fit, it may be challenging to test the Kerr metric with future X-ray data without additional constraints \citep[e.g., study of the timing properties, see][]{2019MNRAS.488..324I}. Fig.~\ref{a13an} shows the $\chi^2$ confidence level curves on the plane $\alpha_{13}$ vs. $\alpha_n$ when we fit the simulated data with {\tt relxilldgrad\_nk} and we do not see any clear correlation between the two parameters. For a conclusive result on the possibility of testing the Kerr metric with \textsl{Athena} and \textsl{eXTP}, we presumably need a full model that has a density gradient and has $E_{\rm cut}$ as a free parameter.

\vspace{0.5cm}

{\bf Acknowledgments --}
This work was supported by the Innovation Program of the Shanghai Municipal Education Commission, Grant No.~2019-01-07-00-07-E00035, the National Natural Science Foundation of China (NSFC), Grant No.~11973019, and Fudan University, Grant No.~JIH1512604. D.A. is supported by a Teach@T{\"ubingen} Fellowship.


\appendix

\section{Relation between $\epsilon$ and $F_X$}

For simplicity, we assume that the corona is point-like and that the photon number flux at the emission point in the corona is described by a power-law
\be\label{eq-corona}
\frac{dN_{\rm c}}{dt_{\rm c} dE_{\rm c}} = \left\{
\begin{array}{cl}
K E^{-\Gamma}_{\rm c} & \quad E_{\rm min} < E < E_{\rm max} \\
0 & \quad \text{otherwise}
\end{array} \right. ,
\ee
where $K$ is a constant, $\Gamma$ is the photon index, and the subindex c reminds that these quantities are evaluated at the emission point in the corona. The relation $F_X \propto g^{2-\Gamma} \epsilon$, where $g$ is the redshift experienced by photons when they travel from the corona to the disk, results from the fact that $\epsilon$ is defined as a specific flux while $F_X$ is a flux integrated over an energy range.

If we think of calculating numerically $\epsilon$ and $F_X$, we can divide the accretion disk into annuli (radial bins). The annulus $i$ will have radial coordinate $r_i$ and width $\Delta r_i$. We can then fire photons from the point-like corona to the disk and infer the ray number per radial bin in the disk, say $\mathcal{N}_i$. The ray number density of the radial bin $i$ is
\be
n_i = \frac{\mathcal{N}_i}{A_i \gamma_i} \, ,
\ee
where $A_i$ and $\gamma_i$ are, respectively, the area of the radial bin $i$ and the Lorentz factor of the particles in the radial bin $i$ measured by the distant observer. The photon number is conserved, so $N = K E^{-\Gamma} \, \Delta t \, \Delta E$ is a constant along the photon path and can be associated to the number of photons for every ray. The energy density illuminating the disk at the radial bin $i$ is
\be\label{eq-ed}
\mathcal{E}_i = E_{\rm d} \, N \, n_i 
= E_{\rm d} \left( K E^{-\Gamma}_{\rm c} \, \Delta t_{\rm c} \, \Delta E_{\rm c} \right) n_i
= g^\Gamma_i \, K E^{-\Gamma+1}_{\rm d} \, \Delta t_{\rm d} \, \Delta E_{\rm d} \, n_i \, ,
\ee
where $g_i = \Delta t_{\rm c}/\Delta t_{\rm d} = E_{\rm d}/E_{\rm c}$ is the redshift factor of the radial bin $i$ between the emission point in the corona and the incident point in the disk and the subindex d is used for the quantities on the disk. The emissivity profile of the radial bin $i$ can be written as
\be
\varepsilon_i \propto \frac{\mathcal{E}_i}{\Delta t_{\rm d} \, \Delta E_{\rm d}}
= g^\Gamma_i \, K E^{-\Gamma+1}_{\rm d} \, n_i \, .
\ee

For the flux illuminating the disk, $F_X$, we can proceed as above and get the energy density illuminating the disk at the radial bin $i$, Eq.~(\ref{eq-ed}). If we integrate over the energy
 \be\label{eq-cc}
 \int_{E_1}^{E_2} g^\Gamma_i \, K E^{-\Gamma+1}_{\rm d} \, \Delta t_{\rm d} \, d E_{\rm d} \, n_i
 = \frac{g^\Gamma_i \, K \Delta t_{\rm d} \, n_i}{2 - \Gamma} \left[ E_2^{2 - \Gamma} - E_1^{2 - \Gamma} \right] \, .
 \ee
Note that $E_1 = g_i E_{\rm min}$ and $E_2 = g_i E_{\rm max}$ and Eq.~(\ref{eq-cc}) becomes
 \be\label{eq-fx}
 \int_{E_1}^{E_2} g^\Gamma_i \, K E^{-\Gamma+1}_{\rm d} \, \Delta t_{\rm d} \, d E_{\rm d} \, n_i
 = \frac{g^2_i \, K \Delta t_{\rm d} \, n_i}{2 - \Gamma} \left[ E_{\rm max}^{2 - \Gamma} - E_{\rm min}^{2 - \Gamma} \right] \, .
 \ee 
  For the flux $F_X$, we divide the above expression by $ \Delta t_{\rm d}$ and we find (omitting the subindex $i$)
 \be
F_X \propto g^{2 - \Gamma} \, \varepsilon \, .
\ee



\begin{thebibliography}{99}

\bibitem[Arnaud(1996)]{1996ASPC..101...17A} Arnaud, K.~A.\ 1996, Astronomical Data Analysis Software and Systems V, 101, 17

\bibitem[Abdikamalov et al.(2019)]{2019ApJ...878...91A} Abdikamalov, A.~B., Ayzenberg, D., Bambi, C., et al.\ 2019, \apj, 878, 91. doi:10.3847/1538-4357/ab1f89

\bibitem[Abdikamalov et al.(2020)]{2020ApJ...899...80A} Abdikamalov, A.~B., Ayzenberg, D., Bambi, C., et al.\ 2020, \apj, 899, 80. doi:10.3847/1538-4357/aba625

\bibitem[Abdikamalov et al.(2021)]{2021PhRvD.103j3023A} Abdikamalov, A.~B., Ayzenberg, D., Bambi, C., et al.\ 2021, \prd, 103, 103023. doi:10.1103/PhysRevD.103.103023

\bibitem[Bambi(2017a)]{2017RvMP...89b5001B} Bambi, C.\ 2017a, Reviews of Modern Physics, 89, 025001. doi:10.1103/RevModPhys.89.025001

\bibitem[Bambi(2017b)]{2017bhlt.book.....B} Bambi, C.\ 2017b, Black Holes: A Laboratory for Testing Strong Gravity, ISBN 978-981-10-4523-3. Springer Nature Singapore Pte Ltd., 2017b. doi:10.1007/978-981-10-4524-0

\bibitem[Bambi et al.(2017)]{2017ApJ...842...76B} Bambi, C., C{\'a}rdenas-Avenda{\~n}o, A., Dauser, T., et al.\ 2017, \apj, 842, 76. doi:10.3847/1538-4357/aa74c0

\bibitem[Bambi et al.(2021)]{2020arXiv201104792B} Bambi, C., Brenneman, L.~W., Dauser, T., et al.\ 2021, \ssr, 217, 65. doi:10.1007/s11214-021-00841-8

\bibitem[Brenneman(2013)]{2013mams.book.....B} Brenneman, L.\ 2013, Measuring the Angular Momentum of Supermassive Black Holes, SpringerBriefs in Astronomy. ISBN 978-1-4614-7770-9. Laura Brenneman, 2013. doi:10.1007/978-1-4614-7771-6

\bibitem[Cao et al.(2018)]{2018PhRvL.120e1101C} Cao, Z., Nampalliwar, S., Bambi, C., et al.\ 2018, \prl, 120, 051101. doi:10.1103/PhysRevLett.120.051101

\bibitem[Dauser et al.(2010)]{2010MNRAS.409.1534D} Dauser, T., Wilms, J., Reynolds, C.~S., et al.\ 2010, \mnras, 409, 1534. doi:10.1111/j.1365-2966.2010.17393.x

\bibitem[Dauser et al.(2013)]{2013MNRAS.430.1694D} Dauser, T., Garcia, J., Wilms, J., et al.\ 2013, \mnras, 430, 1694. doi:10.1093/mnras/sts710

\bibitem[Dov{\v{c}}iak et al.(2004)]{2004ApJS..153..205D} Dov{\v{c}}iak, M., Karas, V., \& Yaqoob, T.\ 2004, \apjs, 153, 205. doi:10.1086/421115

\bibitem[Draghis et al.(2020)]{2020ApJ...900...78D} Draghis, P.~A., Miller, J.~M., Cackett, E.~M., et al.\ 2020, \apj, 900, 78. doi:10.3847/1538-4357/aba2ec

\bibitem[Fabian et al.(1989)]{1989MNRAS.238..729F} Fabian, A.~C., Rees, M.~J., Stella, L., et al.\ 1989, \mnras, 238, 729. doi:10.1093/mnras/238.3.729

\bibitem[Garc{\'\i}a \& Kallman(2010)]{2010ApJ...718..695G} Garc{\'\i}a, J. \& Kallman, T.~R.\ 2010, \apj, 718, 695. doi:10.1088/0004-637X/718/2/695

\bibitem[Garc{\'\i}a et al.(2013)]{2013ApJ...768..146G} Garc{\'\i}a, J., Dauser, T., Reynolds, C.~S., et al.\ 2013, \apj, 768, 146. doi:10.1088/0004-637X/768/2/146

\bibitem[Garc{\'\i}a et al.(2014)]{2014ApJ...782...76G} Garc{\'\i}a, J., Dauser, T., Lohfink, A., et al.\ 2014, \apj, 782, 76. doi:10.1088/0004-637X/782/2/76

\bibitem[Ingram et al.(2019)]{2019MNRAS.488..324I} Ingram, A., Mastroserio, G., Dauser, T., et al.\ 2019, \mnras, 488, 324. doi:10.1093/mnras/stz1720

\bibitem[Johannsen(2013)]{2013PhRvD..88d4002J} Johannsen, T.\ 2013, \prd, 88, 044002. doi:10.1103/PhysRevD.88.044002

\bibitem[Kammoun et al.(2019)]{2019MNRAS.485..239K} Kammoun, E.~S., Dom{\v{c}}ek, V., Svoboda, J., et al.\ 2019, \mnras, 485, 239. doi:10.1093/mnras/stz408

\bibitem[Laor(1991)]{1991ApJ...376...90L} Laor, A.\ 1991, \apj, 376, 90. doi:10.1086/170257

\bibitem[Mastroserio et al.(2021)]{2021MNRAS.507...55M} Mastroserio, G., Ingram, A., Wang, J., et al.\ 2021, \mnras, 507, 55. doi:10.1093/mnras/stab2056

\bibitem[Mitsuda et al.(1984)]{1984PASJ...36..741M} Mitsuda, K., Inoue, H., Koyama, K., et al.\ 1984, \pasj, 36, 741

\bibitem[Nandra et al.(2013)]{2013arXiv1306.2307N} Nandra, K., Barret, D., Barcons, X., et al.\ 2013, arXiv:1306.2307

\bibitem[Negoro et al.(2019)]{2019ATel12968....1N} Negoro, H., Nakajima, M., Sugita, S., et al.\ 2019, The Astronomer's Telegram, 12968

\bibitem[Nied{\'z}wiecki \& {\.Z}ycki(2008)]{2008MNRAS.386..759N} Nied{\'z}wiecki, A. \& {\.Z}ycki, P.~T.\ 2008, \mnras, 386, 759. doi:10.1111/j.1365-2966.2008.12735.x

\bibitem[Nied{\'z}wiecki et al.(2019)]{2019MNRAS.485.2942N} Nied{\'z}wiecki, A., Szanecki, M., \& Zdziarski, A.~A.\ 2019, \mnras, 485, 2942. doi:10.1093/mnras/stz487

\bibitem[Parmar \& White(1985)]{1985IAUC.4051....1P} Parmar, A.~N. \& White, N.~E.\ 1985, \iaucirc, 4051

\bibitem[Parmar et al.(1993)]{1993A&A...279..179P} Parmar, A.~N., Angelini, L., Roche, P., et al.\ 1993, \aap, 279, 179

\bibitem[Reynolds(2014)]{2014SSRv..183..277R} Reynolds, C.~S.\ 2014, \ssr, 183, 277. doi:10.1007/s11214-013-0006-6

\bibitem[Reynolds \& Nowak(2003)]{2003PhR...377..389R} Reynolds, C.~S. \& Nowak, M.~A.\ 2003, \physrep, 377, 389. doi:10.1016/S0370-1573(02)00584-7

\bibitem[Riaz et al.(2020)]{2020arXiv201207469R} Riaz, S., Abdikamalov, A.~B., Ayzenberg, D., et al.\ 2020, arXiv:2012.07469

\bibitem[Risaliti et al.(2013)]{2013Natur.494..449R} Risaliti, G., Harrison, F.~A., Madsen, K.~K., et al.\ 2013, \nat, 494, 449. doi:10.1038/nature11938

\bibitem[Ross \& Fabian(2005)]{2005MNRAS.358..211R} Ross, R.~R. \& Fabian, A.~C.\ 2005, \mnras, 358, 211. doi:10.1111/j.1365-2966.2005.08797.x

\bibitem[Shreeram \& Ingram(2020)]{2020MNRAS.492..405S} Shreeram, S. \& Ingram, A.\ 2020, \mnras, 492, 405. doi:10.1093/mnras/stz3455

\bibitem[Svoboda et al.(2012)]{2012A&A...545A.106S} Svoboda, J., Dov{\v{c}}iak, M., Goosmann, R.~W., et al.\ 2012, \aap, 545, A106. doi:10.1051/0004-6361/201219701

\bibitem[Tripathi et al.(2019a)]{2019ApJ...874..135T} Tripathi, A., Yan, J., Yang, Y., et al.\ 2019a, \apj, 874, 135. doi:10.3847/1538-4357/ab0a00

\bibitem[Tripathi et al.(2019b)]{2019ApJ...875...56T} Tripathi, A., Nampalliwar, S., Abdikamalov, A.~B., et al.\ 2019b, \apj, 875, 56. doi:10.3847/1538-4357/ab0e7e

\bibitem[Tripathi et al.(2020)]{2020MNRAS.498.3565T} Tripathi, A., Liu, H., \& Bambi, C.\ 2020, \mnras, 498, 3565. doi:10.1093/mnras/staa2618

\bibitem[Tripathi et al.(2021a)]{2021ApJ...907...31T} Tripathi, A., Abdikamalov, A.~B., Ayzenberg, D., et al.\ 2021a, \apj, 907, 31. doi:10.3847/1538-4357/abccbd

\bibitem[Tripathi et al.(2021b)]{2021ApJ...913...79T} Tripathi, A., Zhang, Y., Abdikamalov, A.~B., et al.\ 2021b, \apj, 913, 79. doi:10.3847/1538-4357/abf6cd

\bibitem[Tripathi et al.(2021c)]{2021arXiv210610982T} Tripathi, A., Abdikamalov, A.~B., Ayzenberg, D., et al.\ 2021c, arXiv:2106.10982

\bibitem[Verner et al.(1996)]{1996ApJ...465..487V} Verner, D.~A., Ferland, G.~J., Korista, K.~T., et al.\ 1996, \apj, 465, 487. doi:10.1086/177435

\bibitem[Wilms et al.(2000)]{2000ApJ...542..914W} Wilms, J., Allen, A., \& McCray, R.\ 2000, \apj, 542, 914. doi:10.1086/317016 

\bibitem[Zhang et al.(1994)]{1994IAUC.6096....1Z} Zhang, S.~N., Harmon, B.~A., Wilson, C.~A., et al.\ 1994, \iaucirc, 6096

\bibitem[Zhang et al.(2016)]{2016SPIE.9905E..1QZ} Zhang, S.~N., Feroci, M., Santangelo, A., et al.\ 2016, \procspie, 9905, 99051Q. doi:10.1117/12.2232034

\bibitem[Zhang et al.(2019)]{2019ApJ...884..147Z} Zhang, Y., Abdikamalov, A.~B., Ayzenberg, D., et al.\ 2019, \apj, 884, 147. doi:10.3847/1538-4357/ab4271

\end{thebibliography}
\end{document}